\documentclass[aol,twocolumn,showpacs,preprintnumbers,superscriptaddress]{revtex4-1}
\usepackage[bookmarks, colorlinks=true, plainpages = true,linkcolor = blue, citecolor = blue, urlcolor = blue, filecolor =blue]{hyperref}

\usepackage{amsfonts}
\usepackage{graphicx}
\usepackage{amsmath}
\usepackage{amssymb}
\usepackage{latexsym}
\usepackage{psfrag}
\usepackage{dcolumn}
\usepackage{datetime}
\usepackage{multirow}
\usepackage{subfigure}

\begin{document}
\title{Voltage-controlled spin injection with an endohedral fullerene Co@C$_{60}$ dimer}
\author{Alireza Saffarzadeh}
\altaffiliation{Author to whom correspondence should be addressed. Electronic mail: asaffarz@sfu.ca}
\affiliation{Department of Physics, Payame Noor University, P.O.
Box 19395-3697 Tehran, Iran} \affiliation{Department of Physics,
Simon Fraser University, Burnaby, British Columbia, Canada V5A
1S6}
\author{George Kirczenow}
\affiliation{Department of Physics, Simon Fraser University,
Burnaby, British Columbia, Canada V5A 1S6}
\date{\today}

\begin{abstract}
Spin-dependent transport through an endohedral fullerene
Co@C$_{60}$ dimer with gold electrodes is explored theoretically
using density functional and extended H\"{u}ckel theory. Density
of states spin polarizations up to 95\%, due to spin-splitting of
Co $3d$ orbitals, are found by varying the gate and/or bias
voltage. The current-voltage characteristics and strong (up to
100\%) spin polarization of the current indicate that the device
can be utilized for highly efficient spin injection into
nonmagnetic conductors. This finding opens the way to the
realization of electrostatically tuned spintronic nano devices
less than 2 nanometers in size, without ferromagnetic electrodes.
\end{abstract}
\maketitle

Carbon-based nanostructures such as fullerenes, carbon nanotubes
(CNTs), and graphene, are promising candidates for spintronic
applications because of their weak spin-orbit coupling and
hyperfine interaction which lead to long spin coherence lengths
\cite{Rocha,Sanvito}. In particular, the fullerene C$_{60}$ molecule
is an interesting carbon nanostructure that can be used as a
molecular bridge in magnetic tunnel junctions due to its
remarkable structural stability and electronic properties
\cite{Braun,Koleini}.

One way to generate a spin-polarized charge current is to
encapsulate magnetic atoms or magnetic nanowires in fullerenes or
CNTs
\cite{Yang,Yagi,Zare,Fujima,Kang,Sakai1,Sakai2,Miwa,Ivanovskaya,Wang,Parq,Sugai,Xie,Salas,Lu,Sabirianov}.
For example, \textit{ab initio} calculations showed that Co atoms
in CNTs can provide strong spin polarization and considerable
magnetic moments \cite{Yang}. The electrical and magnetic
properties of C$_{60}$-Co nanocomposites have also been studied
\cite{Zare,Sakai2,Miwa} and tunnel magnetoresistance ratios up to
about 30\% at low bias voltages were reported, indicating
significant spin polarizations \cite{Zare}. Recent density
functional theory (DFT) studies of encapsulated Co atoms in
C$_{60}$ molecules \cite{Salas} found hybridization between Co and
C$_{60}$ orbitals and the most stable structure of Co@C$_{60}$ to
have the Co atom on top of a hexagonal face. Lu \textit{et al.}
\cite{Lu} showed that, in the most stable structure of
Gd@C$_{60}$, the Gd ion is over a hexagonal ring of the C$_{60}$
molecule and the Gd atomic orbitals to hybridize with the C$_{60}$
molecular orbitals. Moreover, strong hybridization between the Gd
$5d$ and $6s$ and carbon orbitals in Gd@C$_{60}$ has been
observed, both theoretically and experimentally \cite
{Sabirianov}.

Magnetic atoms encapsulated in carbon nanocages, like the C$_{60}$
molecule, are effectively protected against environmental effects
such as oxidization, stabilizing the encapsulated magnetic atoms
for potential applications for spin injection in nanoscale
devices. Accordingly, one can propose magnetic nano junctions to
produce high spin-polarized currents with long spin coherence
lengths \cite{Tsukagoshi}. In this paper we explore theoretically
the possibility of such molecular magnetic junctions in which the
carbon nanocage is a fullerene dimer. Since, among various types
of fullerene dimers, C$_{120}$ in which the two C$_{60}$ molecules
are connected by a cyclic C$_4$ unit, is the simplest one and has
the lowest energy as well as interesting physical and chemical
properties \cite{Segura}, we have chosen this isomer of the
C$_{60}$ dimer in our calculation. We note that, the electronic
transport properties of a pure C$_{60}$ dimer \cite{He}, N and B
doped C$_{60}$ dimer \cite{Zheng}, and Li@C$_{60}$ dimer
\cite{Zhao} have been reported theoretically, however,
encapsulation of magnetic atoms in fullerene dimers has not been
addressed in the previous theories.

\begin{figure}[t!]
\includegraphics[width=0.95\linewidth]{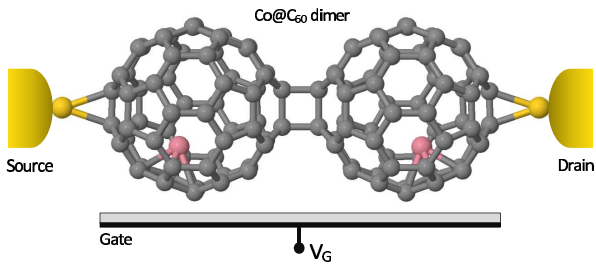}
\caption{(Color online) The relaxed geometry of an endohedral
fullerene Co@C$_{60}$ dimer junction. The gray, pink, and yellow
spheres represent C, Co, and Au atoms, respectively. The effect of
the gate voltage, $V_G$, is to shift the molecular orbitals
relative to the Fermi level of the gold
electrodes.}\label{Figure1}
\end{figure}

We consider an endohedral fullerene Co@C$_{60}$ dimer contacted
via single gold atoms (as in some experimentally realized
single-molecule electronic devices \cite{George, Demir}) by
semi-infinite non-magnetic electrodes, as shown in Fig.
\ref{Figure1}. The electronic density of states, degree of spin
polarization, spin-polarized charge currents and differential
conductances for electron tunneling through
Au/Co@C$_{60}$-dimer/Au junction as a function of bias and gate
voltage in the dimer ground state, which we find to be
ferromagnetic, are studied by means of DFT calculations of the
relaxed geometry and the extended H\"{u}ckel model of quantum
chemistry \cite{George,Ammeter,Yaehmop}. The extended H\"{u}ckel
theory (EHT) has been used to study spin transport in non-magnetic
molecules contacted by magnetic electrodes
\cite{Emberly0,Dalgleish1,Dalgleish2} and in single molecule
magnets coupled to the non-magnetic electrodes \cite{F1,F2}.
However, the EHT does not by itself take account of spin
polarization effects, and for this purpose we generalized the
theory to include spin splittings calculated within DFT.

We carried out \textit{ab initio} DFT geometry relaxations for the
Co@C$_{60}$ dimer sandwiched between two Au atoms using the
GAUSSIAN 09 package with the B3PW91 functional and the Lanl2DZ
basis \cite{Gaussian,Perdew}. In the relaxed structure (Fig.
\ref{Figure1}) the Co atoms are at off-center positions in the
C$_{60}$ molecules, with a shortest Co-C bond length of 2.09
{\AA}. We found the molecules to couple to each other by parallel
bonds with a length of 1.59 {\AA} in agreement with previous
studies \cite{Zheng,Zhao,Trave}. The Au-C and Co-Co distances are
2.26 {\AA} and 10.49 {\AA}, respectively. In Fig. \ref{Figure1},
the distance between leftmost and rightmost carbon atoms is 16.31
{\AA}. Thus our proposed spin-injection device has a size less
than 2 nanometers.

The dimer system in the presence of gold electrodes is described
by a tight-binding model Hamiltonian derived from EHT in a basis
of atomic valence orbitals, the $3d$, $4s$, and $4p$ valence
orbitals of the Co atoms and the $2s$ and $2p$ orbitals of the
carbon atoms. These 498 atomic orbitals form the basis set for the
Co@C$_{60}$ dimer Hamiltonian that can be written for an electron
with spin $\sigma$ as
\begin{equation}\label{01}
H_{D,\sigma}=\sum_{\alpha}\epsilon_{\alpha\sigma}
d_{\alpha\sigma}^\dag d_{\alpha\sigma}+
\sum_{\alpha,\beta}\gamma_{\alpha
\beta,\sigma}(d_{\alpha\sigma}^\dag
d_{\beta\sigma}+\mathrm{h.c.})\  ,
\end{equation}
where $d_{\alpha\sigma}^\dag$ is the creation operator for an
electron in an atomic valence orbital $\psi_{\alpha\sigma}$ with
energy eigenvalue $\epsilon_{\alpha\sigma}$.
$\gamma_{\alpha\beta,\sigma}$ is the spin-dependent matrix element
of the extended H\"{u}ckel Hamiltonian between valence orbitals
$\psi_{\alpha\sigma}$ and $\psi_{\beta\sigma}$. Since the basis
set used in EHT is nonorthogonal, the orbital overlap $S_{\alpha
\beta,\sigma}=\langle\psi_{\alpha\sigma}|\psi_{\beta\sigma}\rangle$
that we obtain from EHT can be non-zero. Therefore, we replace
$\gamma_{\alpha\beta,\sigma}$ in the Hamiltonian by
$\gamma_{\alpha\beta,\sigma}-\epsilon S_{\alpha \beta,\sigma}$
where $\epsilon$ is the electron energy \cite{Emberly1,Emberly2}.
The spin-dependence of $\epsilon_{\alpha\sigma}$ and
$\gamma_{\alpha\beta,\sigma}$ results from the spin splittings of
the $3d$ and $4s$ orbitals of the Co atoms that are estimated from
DFT calculations for major peaks in the spin density of states for
Co on graphene \cite{Chan,Saffar1}. We note that the
spin-dependent electron scattering that arises from these exchange
related spin-splittings is much stronger than that due to
spin-orbit coupling. Therefore we neglect the latter in the
present model. The spin splitting values modify the Co $4s$ and
$3d$ orbitals on-site energies obtained from the EHT parameters.
While the limitations of DFT are well known \cite{George}, the
present theoretical approach has explained \cite{Saffar1} the
results of scanning tunneling spectroscopy experiments \cite{Brar}
on Co atoms adsorbed on graphene. Based on the Hamiltonian Eq.
(\ref{01}), the total Hamiltonian of the Co@C$_{60}$ dimer coupled
to the left ($L$) and right ($R$) gold electrodes for an electron
with spin $\sigma$ can be written as
$H_\sigma=H_{D,\sigma}+\Sigma_{L,\sigma}+\Sigma_{R,\sigma}$. Here,
$\Sigma_{\eta,\sigma}(\epsilon)=\tau_{\eta,\sigma}
g_{\eta}(\epsilon)\tau^\dag_{\eta,\sigma}$ is a self-energy that
describes the coupling of the dimer to the electrode $\eta$ (=$L$
or $R$), $\tau_{\eta,\sigma}$ is the hopping matrix between
electrode $\eta$ and the dimer whose matrix elements are given by
the EHT parameters. For Au atoms in the electrodes we use $6s$,
$6p_x$, $6p_y$, $6p_z$, $5d_{x^2-y^2}$, $5d_{xy}$,
$5d_{3z^2-r^2}$, $5d_{zx}$, $5d_{zy}$ atomic valence orbitals. Due
to the nine valence orbitals for each Au atom, we model the
electrodes, as in previous work \cite{Dalgleish1,Dalgleish2,
F1,F2,Demir,Cardamone}, by nine semi-infinite one-dimensional
chains of orthogonal atomic orbitals, with one orbital per site
and periodicity $a$ between orbital sites. The one-dimensional
chains are decoupled from each other, i.e., there is no hopping
between orbitals in different chains. Accordingly, the surface
Green's function matrix for electrode $\eta$ is
$g_\eta=-(1/t)\,e^{ika}I$ where $k$ is the electron wave number
and $I$ is a $9\times 9$ unit matrix \cite{Datta}. We note that,
the overlap matrix $S$ between the dimer and Au orbitals is
included in the calculations. In this study, the Co $3d$ orbitals
have been shifted in energy so that the average electric charge,
corresponding to the two spin-up and spin-down states of each Co
atom in the antiferromagnetic alignment and calculated from Eq.
\ref{01}, is close to the Mulliken charge obtained from DFT using
the GAUSSIAN 09 package. In addition, to locate the position of
Fermi energy of the gold electrodes between the HOMO (highest
occupied molecular orbitals) and LUMO (lowest unoccupied molecular
orbital) energies of the Co@C$_{60}$ dimer, all the atomic valence
orbitals of the dimer were shifted down in energy by $0.4\,t$ (in
units of $t=2.7$ eV), relative to their energies in the extended
H\"{u}ckel model.

The magnetic moments of the two Co atoms can be parallel
(ferromagnetic (FM)) or antiparallel (antiferromagnetic (AF)). In
the FM case, a spin-up electron travels from a majority spin state
of one C$_{60}$ molecule, induced by its Co atom, to a majority
spin state of the other C$_{60}$ molecule, induced by the other Co
atom, while spin-down electrons travel from a minority spin to a
minority spin state. In the AF alignment, electrons belonging to
the majority spin state of one molecule, travel to the minority
spin states of the other molecule and visa versa. To find the
magnetic ground state configuration, we estimate the total ground
state energy for each spin state in the magnetic alignment $\xi$
(=FM or AF) as $E_{\mathrm{total},\sigma}^{\,\xi}=\sum_{i}
E_{i,\sigma}^{\,\xi}$ where $E_{i,\sigma}^{\,\xi}$ is the $i$-th
molecular orbital eigenenergy for $\sigma$ spin state and the
summation is over occupied molecular orbitals. We found the lowest
total energy to correspond to the spin-up state of the FM
alignment whose energy is lower than for the AF alignment by $\sim
5.24$ eV. Accordingly, to switch the Co@C$_{60}$ dimer from FM to
AF alignment, a strong magnetic field is required. In the absence
of a strong field, the Co@C$_{60}$ dimer will remain permanently
in its FM state. In the present system the strong hybridization of
the Co 3$d$ orbitals and carbon valence states results in indirect
exchange between the Co atoms (mediated by the carbon valence
orbitals) that stabilizes the FM configuration that has also been
reported in analogous structures \cite{Xu,Xiao,Robles}. Therefore,
to study the effect of encapsulated Co atoms on electronic
transport through the dimer bridge, only the FM alignment will be
considered.

\begin{figure}[t!]
\includegraphics[width=0.7\linewidth]{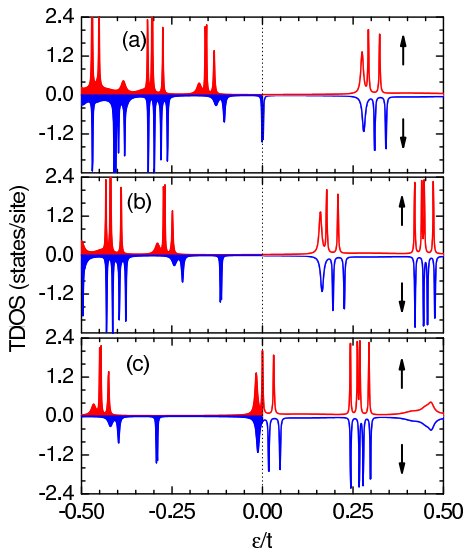}
\caption{(Color online) Calculated carbon atoms TDOS per site vs.
electron energy $\epsilon$ for the endohedral fullerene
Co@C$_{60}$ dimer at various gate voltages: (a) $V_G=0.115\,t/e$,
(b) $V_G=0.0\,t/e$, (c) $ V_G=-0.177\,t/e$. The symbol $\uparrow$
($\downarrow$) corresponds to spin-up (spin-down) electrons. The
vertical dotted lines indicate the position of the Fermi level.
All states up to the Fermi level are occupied (shaded in the
plot).}\label{Figure2}
\end{figure}
\begin{figure}[t!]
\includegraphics[width=0.7\linewidth]{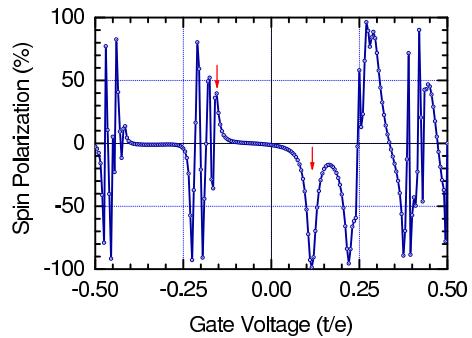}
\caption{(Color online) Degree of density of sates spin
polarization $P$ at the Fermi energy as a function of gate
voltage. The arrows show the positions of the first maxima of spin
polarization with varying gate voltage.}\label{Figure3}
\end{figure}

To calculate the spin states and the degree of spin polarization
in this system, we define the spin-dependent Green's function of
the Co@C$_{60}$ dimer coupled to the gold electrodes as
$G_{\sigma}(\epsilon)=[(\epsilon+i\delta)S_{\sigma}-H_{\sigma}]^{-1}$
where $\delta$ is a positive infinitesimal. Then the total density
of states (TDOS) of carbon atoms per site for spin $\sigma$ is
$D_{\sigma}(\epsilon)=-\frac{1}{N\pi}\sum^{4N}_\alpha\mathrm{Im}(\langle
\alpha,\sigma|G_{\sigma}S_{\sigma}|\alpha,\sigma\rangle)$, where
the sum is over the carbon $2s$, $2p_x$, $2p_y$, and $2p_z$
orbitals, and $N=120$ is the number of carbon atoms in the dimer.
The degree of density of states spin polarization $P$ is defined
as
$P=(D_\uparrow(E_F)-D_\downarrow(E_F))/(D_\uparrow(E_F)+D_\downarrow(E_F))$,
where $D_\uparrow(E_F)(D_\downarrow(E_F))$ represents the TDOS of
spin-up (spin-down) electrons at the Fermi energy. Note that, the
spin dependence of electronic states at the sites of carbon atoms
originates from hybridization between carbon orbitals and Co 4$s$
and 3$d$ orbitals. To see the effect of this hybridization on the
electronic states, we show in Fig. \ref{Figure2} the TDOS as a
function of electron energy for representative gate voltages
$V_G=+0.115$, 0.0, and $-0.177\,t/e$. In our calculations, the
extended H\"{u}ckel model parameters describing both the on-site
and intersite Hamiltonian matrix elements are recalculated for
each value of the gate voltage. The dependence of the intersite
Hamiltonian matrix elements on the gate voltage (see Ref. 30,
Appendix A) implies that hybridization between the orbitals on
different atoms depends on the gate voltage in the present theory.
However, since the same gate potential is applied to the C$_{60}$
molecules and Co atoms, the molecular orbital features shift
nearly rigidly under gating. The chosen positive and negative gate
voltage values in Fig. \ref{Figure2} correspond to the first peaks
(in the absolute value of $P$ in the positive and negative region
of gate voltage) crossing the Fermi level (see the arrows in Fig.
\ref{Figure3}). At zero gate voltage the Fermi level is located in
the HOMO-LUMO gap where no molecular orbital is available for
electrons to tunnel through the dimer (see Fig. \ref{Figure2}(b)),
and hence little or no spin polarization is expected. By applying
a gate voltage, the encapsulated Co atoms are ionized due to the
gate-induced shift in energy of the molecular orbitals relative to
the gold electrode's Fermi energy. The effects of the positive and
the negative gate voltage, have been shown in Fig.
\ref{Figure2}(a) and 2(c), respectively. Comparing Fig.
\ref{Figure2}(b) with Fig. \ref{Figure2}(a) and 2(c), it is clear
that a positive (negative) gate voltage shifts the molecular
orbitals to a higher (lower) energy and for the value of
$V_G=+0.115\,t/e$ at which the spin down states are conductive,
there is a significant difference between the two spin populations
which can cause a strong spin polarization at the Fermi level as
shown in Fig. \ref{Figure3}. On the other hand, in the case of
$V_G=-0.177\,t/e$ the molecular orbitals shift to a lower energy
and, as shown in Fig. \ref{Figure3}, a peak in the spin
polarization appears.
The results shown in Fig. \ref{Figure3} for $P$ versus $V_G$
indicate that the spin injection can be controlled by changing the
gate voltage and positive or negative values for the spin
polarization may be obtained, depending on the value and the sign
of gate potential. The magnitude of $P$ can exceed 90\% for
certain values of gate voltage which is considerably higher than
that calculated for Co-nanowire@CNT (66\%) \cite{Yang} and those
measured for Co-cluster/C$_{60}$ mixtures (70\%) \cite{Zare},
Co-nanoparticle/Co-C$_{60}$ compounds (40\%) \cite{Sakai2} and Co
crystals (42\%) \cite{Soulen}.

To investigate coherent charge transport through the Co@C$_{60}$
dimer, we make use of the Landauer-B\"{u}ttiker formula based on
the nonequilibrium Green's function method \cite{Datta} for which
the spin-polarized charge current at a constant bias voltage,
$V_b$, is calculated as
\begin{equation}\label{IV}
I_\sigma(V_b)=
\frac{e}{h}\int_{-\infty}^{\infty}T_\sigma(\epsilon,V_b)
[f(\epsilon-\mu_L)-f(\epsilon-\mu_R)]d\epsilon\  ,
\end{equation}
where $f(\epsilon)$ is the Fermi function,
$\mu_{L,R}=\epsilon_F\pm\frac{1}{2}eV_b$ are the chemical
potentials of the electrodes,
$T_\sigma=\mathrm{Tr}[\Gamma_{L,\sigma}G_{\sigma}\Gamma_{R,\sigma}G^\dag_\sigma]$
is the spin-dependent transmission function for electron tunneling
through the dimer, and
$\Gamma_{\eta,\sigma}=-2\,\mathrm{Im}(\Sigma_{\eta,\sigma})$ is
the coupling matrix between the dimer and the electrode $\eta$.

\begin{figure}[t!]
\includegraphics[width=0.9\linewidth]{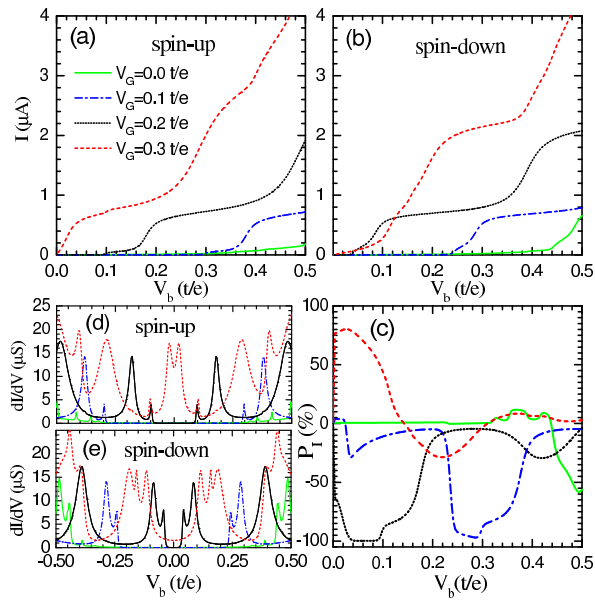}
\caption{(Color online) Calculated spin-dependent electric
currents [(a), (b)], current spin polarization $P_I$ (c), and
differential conductances [(d), (e)] vs bias voltage for the
Au/Co@C$_{60}$-dimer/Au junction, shown in Fig. \ref{Figure1}, at
various $V_G$ values.} \label{Figure4}
\end{figure}

The calculated current-voltage characteristics, $I$-$V$, and the
differential conductance spectra, d$I$/d$V$, of the
Au/Co@C$_{60}$-dimer/Au junction as functions of bias voltage
$V_b$ at $V_G=0.0$, 0.1, 0.2, and 0.3$\,t/e$ and at temperature
$T=4.2$ K for spin-up and spin-down electrons are shown in Fig.
\ref{Figure4}. Due to the structural symmetry in the system, we
obtained a symmetrical behavior in the $I$-$V$ curves with respect
to $V_b=0.0\,t/e$. Therefore, the results for negative bias
voltages are not shown in Fig. \ref{Figure4}(a) and 4(b). The spin
up and spin down charge currents show a step-like behavior due to
the entry of molecular orbitals into the bias window. The
sharpness of the steps depends on the hybridization of the dimer
electronic states shown in Fig. \ref{Figure2} to the gold
electrode orbitals. The broadening of the quantized states in the
scattering region causes smooth steps in the $I$-$V$ curves for
both spin directions. We see that the $I$-$V$ characteristics of
the Co@C$_{60}$ dimer reveal a conducting or semiconducting
behavior depending on the value of gate and bias voltages. This
electronic property is strongly spin dependent due to the spin
splitting of Co $3d$ orbitals \cite{Chan,Saffar1} around the Fermi
energy of gold electrodes. This is consistent with the work of Xie
\textit{et al.} \cite{Xie} who found that hybridization between Co
$3d$ and C $2p$ orbitals for Co nanowires encapsulated in CNTs is
strong and results in spin polarization that increases with
increasing thickness of the nanowire.

At $V_G=0.0$, 0.1, and $0.2\,t/e$, a threshold voltage that is
different for spin-up and spin-down electrons, is needed to
generate a charge current through the device. That is, at low gate
voltages for low bias voltages the device is in an off state,
while at $V_G=0.3\,t/e$, the device is turned on and the charge
current increases linearly, indicating Ohmic behavior at low
biases. Such a behavior is reasonable because in this situation,
the gate voltage shifts the molecular orbitals within the bias
window and more channels for tunneling electrons through the
Co@C$_{60}$ dimer bridge become accessible. Furthermore, by
comparing the charge currents in Fig. \ref{Figure4}(a) and 4(b) at
each bias voltage, one can readily confirm that the tunneling
channels for spin-up and spin-down electrons are well separated
due to the presence of Co atoms inside the dimer. Thus, our model
predicts gate and bias voltage-controlled switching between off
and on states and also between spin-up and spin-down electric
currents. In addition, the results shown in Fig. \ref{Figure4}(a)
and 4(b) reveal that the encapsulation of Co atoms in fullerene
cages is an efficient way for injecting spin-polarized electrons
into a nonmagnetic conductor by electric current. To estimate the
spin injection efficiency in our model, we calculate the current
spin polarization defined as
$P_I=(I_\uparrow-I_\downarrow)/(I_\uparrow+I_\downarrow)$ for each
bias and gate voltage. The results for $P_I$ as a function of
$V_b$ are shown in the Fig. \ref{Figure4}(c). We clearly see that
very high values for current spin polarization ($\sim 99\%$) can
be obtained in this hybrid structure. These values depend strongly
on the bias and gate voltages and predict that highly efficient
voltage-controlled spin injection can be achieved using the
Co@C$_{60}$ dimer.

The differential conductances shown in Fig. \ref{Figure4}(d) and
(e) clarify the role of gate and bias voltages in controlling the
spin-polarized transport through the system. To show the
conductance gap, i.e., the voltage width of the zero-conductance
region, we have plotted the d$I$/d$V$ spectra for both positive
and negative bias voltages. In Fig. \ref{Figure4}(d) and (e) the
size of the conductance gap decreases and the system gradually
moves into a conducting state with increasing $V_G$. The symmetry
of the differential conductances about $V_b=0$ confirms the
existence of symmetry in the $I$-$V$ curves of Fig.
\ref{Figure4}(a) and (b). The sharp conductance peaks correspond
to the step-like features seen in the $I$-$V$ curves and signal
the opening of new conducting channels through the dimer.

In conclusion, we have shown, based on {\em ab initio} and
semi-empirical calculations, that very high degrees of spin
polarization can be achieved for electrons tunneling through
endohedral fullerene Co@C$_{60}$ dimers in their ferromagnetic
ground state. Our results for spin-dependent $I$-$V$ curves and
d$I$/d$V$ spectra, calculated by means of Landauer-B\"{u}ttiker
theory, show that these hybrid structures can be utilized for
highly efficient spin injection into a nonmagnetic conductor by
adjusting gate and/or bias voltages. This {\em electrostatic}
activation should lend itself to integration of such nano spin
injectors with conventional nano electronic devices. Also, since
magnetic electrodes are not required to achieve spin injection in
this system, the size of the device required for spin injection is
that of the Co@C$_{60}$ dimer itself, i.e., less than 2
nanometers. Therefore as well as being of fundamental interest,
systematic experimental studies of encapsulated magnetic atoms in
the fullerene dimer based on these findings may be relevant for
nanoelectronic and spintronic applications.

This work was supported by NSERC, CIFAR, WestGrid, and Compute
Canada.

\end{document}